\documentclass[aps,superscriptaddress,prd,
onecolumn,
floatfix, 
nofootinbib,
amsmath,amssymb,amsfonts,longbibliography]{revtex4-2}
\usepackage[pdftex]{graphicx}
\usepackage{hyperref}
\hypersetup{
setpagesize=false,
bookmarksnumbered=true,
bookmarksopen=true,%
colorlinks=true,%
linkcolor=red,
citecolor=blue,
}
\usepackage{float,color}
\usepackage{bm}
\usepackage{rsfso}
\usepackage{physics}
\usepackage{comment}

\newcommand{\ep}{\epsilon}

\begin{document}

\title{Second-Order Coherence as an Indicator of Quantum Entanglement of Hawking Radiation in Moving-Mirror Models}
% \anno{[title OK?]}
\author{Masanori Tomonaga }
\email{tomonaga.masanori.b3@s.mail.nagoya-u.ac.jp}
\author{Yasusada Nambu}
\email{nambu.yasusada.e5@f.mail.nagoya-u.ac.jp}
%\email{nambu@gravity.phys.nagoya-u.ac.jp}
\affiliation{Department of Physics, Graduate School of Science, Nagoya University, Chikusa, Nagoya 464-8602, Japan}

\date{\today}

\begin{abstract}
%\anno{[reconsider: `intensity correlation' or `second order coherence', too many words `second order coherence']}
  %\anno{Second-order coherence} is well known in the field of quantum optics and is a useful physical quantity for investigating the quantum nature of light. In fact, \anno{second-order coherence} can estimate the quantum entanglement of a light source. In this paper, we investigate the behavior of the \anno{second-order coherence} in a moving mirror system, which is an analog model of Hawking radiation from a black hole. Based on the relation between \anno{second-order coherence} and entanglement of Hawking radiation in the moving mirror model, we discuss how \anno{second-order coherence} can be used as an indicator of quantum entanglement in the entanglement harvesting protocol with two qubit detectors. 
 %\annob{
  The second-order coherence of light is a widely recognized physical quantity used to assess the quantum characteristics of light, and its properties have been extensively investigated in the field of quantum optics. Recently, it has been proposed that second-order coherence can be utilized as an indicator of quantum entanglement. In this study, we evaluated the second-order coherence in the context of the moving-mirror model, which serves as an analog model for Hawking radiation from a black hole. We discuss the relation between entanglement and the second-order coherence of Hawking radiation paying attention to the ``noise" effect due to the thermality of Hawking radiation, which reduces the quantum correlation in the entanglement-harvesting protocol with two-qubit detectors.
  %}
\end{abstract}
\maketitle
%\tableofcontents

%%%%%%%%%%%%%%%%%%%%%%%%%%%%%%%%%%%%%%%%%
%\anno{
%Comment:
%\begin{itemize}
%\item Please check `red' colored part.
%\item 
%\end{itemize}
%}
%%%%%%%%%%%%%%%%%%%%%%%%%%%%%%%%%
\section{Introduction}
Hawking proposed the concept of black hole (BH) evaporation in 1974 \cite{Hawking-1974}. However, this phenomenon causes a serious problem known as the information loss paradox \cite{Raju-2022}. BHs are formed by the gravitational collapse of massive stars. In the Hawking scenario, after the formation of a BH, it evaporates by emitting thermal radiation.  The conceptual difficulty of this scenario is that if we consider the pure quantum state of the BH system at its formation, the final state after evaporation becomes a mixed state. Therefore, this scenario violates the principle of unitarity. The reason for this is probably our lack of understanding of the evaporation process of BHs. To address this issue, many studies have focused on purification partners of Hawking radiation, such as BH remnants \cite{Lin-Hu-2008}, BH soft hairs \cite{Hawking-Perry-Strominger-2016, Hotta-Nambu-Yamaguchi-2018}, and vacuum fluctuations \cite{Wilczek-1993, Hotta-Schutzhold-Unruh-2015}. In either case, understanding the entanglement structure of a BH with Hawking radiation will help to solve this problem.

Recently, to extract entanglement from the quantum field, an entanglement-harvesting protocol using a two-qubit detector system was introduced. Many studies and much research have investigated the structure of quantum entanglement in Hawking radiation \cite{Tjoa-Mann-2021, Yoshimura-Tjoa-Mann-2021, Yosimura-Mann-2021, Liu-Zhang-Yu-2023, Naeem-Yoshimura-Mann-2023, Simidzija-Martin-Martinez-2018, Simidzija-Martinez-2017, Lin-Zhang-Yu-2021}. In this protocol, quantum informational quantities such as negativity, concurrence, and mutual information can be evaluated on the basis of the quantum state tomography technique of the two-detector system. In contrast, many researchers in the field of quantum optics are attempting to obtain the quantum nature of light directly from the response of detectors without utilizing the tomography of the state. Characterized by second-order coherence, the intensity correlation of light serves as an important parameter for assessing the quantum nature. The application of the quantum nature of light is in the field of quantum optics. This is because light with quantum entanglement, which is a key property of the quantum aspect of light, can be applied to subjects such as super-resolution and noise resistance imaging \cite{Kolobov-2007, Bennink-Bentley-Boyd-2002, Shapiro-2008, Lemos-Borish-Cole-Ramelow-Lapkiewicz-Zeilinger-2014}.
Second-order coherence can characterize the quantum nature of light; however,  it has been pointed out in recent years that it is also possible to estimate quantum entanglement \cite{Strobinska-Wodkiewicz-2005, Cotler-Wilczek-Borish-2021}. 
To date, the evaluation of quantum entanglement based on second-order coherence has been limited to only two independent photon modes. 

In this study, we consider how this quantity can be used as a measure of quantum entanglement within the framework of quantum field theory. For this purpose, we investigated its behavior by comparing it with the negativity of entanglement within the framework of the entanglement-harvesting protocol with two-qubit detectors \cite{Clerk-Devoret-Girvin-Marquardt-Schoelkopf-2010}.
We especially pay attention to the behavior of the second-order coherence under the influence of ``thermal noise" originating from Hawking radiation.

The remainder of this paper is organized as follows. In Section \ref{sec. setup}, we present a moving-mirror model and an entanglement harvesting protocol with a two-qubit detector system. In Section \ref{sec. coherence}, we introduce second-order coherence and derive its relationship with entanglement negativity. In Section \ref{sec. result}, we present numerical results and investigate how the ``noise"
%thermal noise (Hawking radiation) 
affects in the harvesting protocol. Section \ref{sec. conclusion} is devoted to the summary and further developments of this study.
%we conclude this paper.

%%%%%%%%%%%%%%%%%%%%%%%%%%%%%%%%%%%%%%%%%%%%%%%%%%%%%%%%%%%%%%%%
%%%%%%%%%%%%%%%%%%%%%%%%%%%%%%%%%%%%%%%%%%%%%%%%%%%%%%%%%%%%%%%%
%%%%%%%%%%%%%%%%%%%%%%%%%%%%%%%%%%%%%%%%%%%%%%%%%%%%%%%%%%%%%%%%

\section{Setup}\label{sec. setup}

\subsection{Moving mirror model}\label{sec. moving mirror}

In this study, several quantities of quantum information were evaluated using the moving-mirror model. The moving-mirror model mimics the evaporation of BHs as the dynamics of a massless scalar field with a moving boundary and captures the intrinsic nature of BH evaporation without a backreaction effect. Therefore, several researchers have studied this model \cite{Carlitz-Willey-1987,Davies-Fulling-1977, Osawa-Lin-Nambu-Hotta-2024}. 

\begin{figure}[t]
  \centering
  \includegraphics[width=0.3\linewidth]{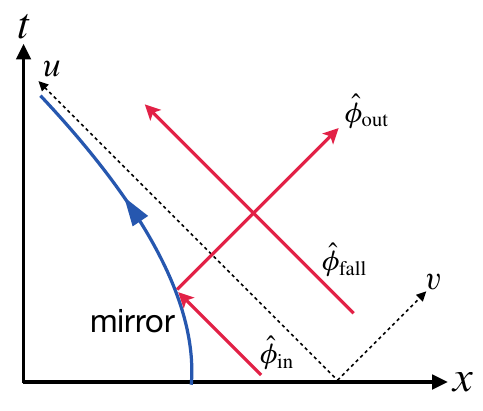}
  \hspace{1cm}
  \includegraphics[height=0.18\linewidth]{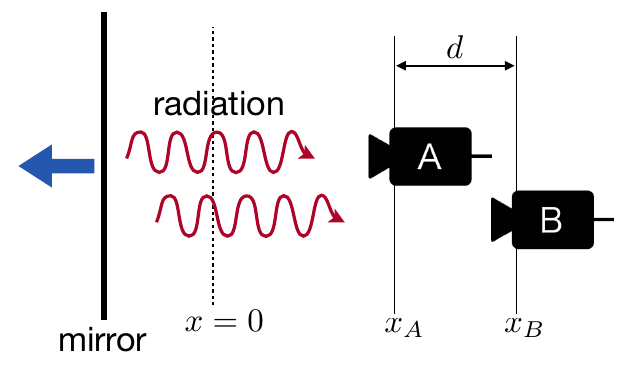}
  \caption{The left panel shows a schematic picture of the moving-mirror model. The blue line denotes the trajectory of the mirror. The incident wave $\hat{\phi}_{\text{in}}$  coming from the $v<0$ region is reflected by the mirror to become $\hat{\phi}_{\text{out}}$. The incident wave $\hat{\phi}_{\text{fall}}$  coming from the $v>0$ region is not reflected by the mirror and does not change its direction. The reflected wave corresponds to analog Hawking radiation.
  The right panel shows the schematic setup of the entanglement-harvesting protocol with the moving-mirror model.}
  \label{fig: mirror model}
\end{figure}

We consider a 1+1 dimensional Minkowski spacetime and introduce null coordinates $u=t-x, v=t+x$. Here, the mirror trajectory is assumed to be $v = p(u)$, where $p(u)$ is a ray-tracing function that characterizes the propagation of the wave modes in the model. The dynamics of a quantized massless scalar field in a flat spacetime is represented by the Klein-Gordon equation (KG)
$\partial_u\partial_v\hat{\phi}(u,v) = 0$.
In the moving-mirror model, the mirror functions as a boundary with perfect reflection. Therefore, in the mirror position, the scalar field must satisfy the Dirichlet boundary condition $\hat{\phi}(t,x_{\text{mirror}})=0$. The field operator can be expanded as:
%%%
\begin{align}\label{eq: field operator}
  \hat{\phi}(u,v) = \int^{\infty}_{0} dk \left[\hat{a}_k \varphi_k(u,v) + \hat{a}_k^{\dagger}\varphi_k^*(u,v)\right],
\end{align}
%%%
where $\varphi_k$ denotes the normalised mode solution to  the KG equation under the mirror boundary condition:  
%%%
\begin{align}\label{eq: mode solution}
  \varphi_k(u,v) = \frac{1}{\sqrt{4\pi k}}\left[e^{-ikv} - e^{-ikp(u)}\right],
\end{align}
%%%
and $\hat{a}_k$ denotes the annihilation operator for the in-vacuum state $\ket{0}_{\text{in}}$. In this study, we assume the following form of the ray-tracing function:
%%%
\begin{align}
  p(u) = -\frac{1}{\kappa} \ln\left(1 + e^{-\kappa u}\right),
  \label{eq: ray-tracing}
\end{align}
%%%
where the constant $\kappa$ denotes the acceleration of the mirror. The trajectory of the mirror is indicated by the blue line in the left panel of Fig.~\ref{fig: mirror model}. The mirror starts a constant acceleration motion with a velocity of zero and then asymptotically approaches a null motion. 
For the mirror trajectory shown in Fig.~\ref{fig: mirror model}, there are two types of incident scalar waves: one reflected by the mirror and the other not. Waves that are not reflected by the mirror represent wave modes that fall into the BH. In the standard scenario of BH evaporation, the fall mode returns to the observable region after completion of the evaporation. Therefore, the existence of unreflected waves corresponds to an scenario of eternal BH, in which the evaporation process will not terminate.
%Hence it corresponds to the eternal black holes.\anno{[why?]}
%\anno{[explain feature of this trajectory]}
%%%

%Under this mirror trajectory, there are two type waves, one reflected by the mirror and another is not (see Fig.\ref{fig: mirror model}). The reflected wave $\phi_{\text{out}}$ is considered analogous to Hawking radiation. Actually, by calculating the vacuum expectation value of the energy flux of the reflected wave $\hat{T}_{uu} = :\left(\partial_u\phi_{\text{out}}\right)^2:$ at $\mathcal{S}^+$, we can confirm $\bra{0_{\text{in}}}\hat{T}_{uu}\ket{0_{\text{in}}} \sim \frac{\pi}{12}T^2$, where $T = \kappa / 2\pi$.

The reflected wave mode corresponds to the analogous Hawking radiation. The expectation value of the energy momentum tensor of the reflected wave is calculated as follows:
%%%
\begin{align}
    \bra{0}_{\text{in}}\hat{T}_{uu}\ket{0}_{\text{in}} = \lim_{\epsilon \to 0}\left[\bra{0}_{\text{in}}\partial_{u}\hat{\phi}_{\text{out}}(u+\epsilon)\partial_{u}\hat{\phi}_{\text{out}}(u)\ket{0}_{\text{in}} - \bra{0}_{\text{out}}\partial_{u}\hat{\phi}_{\text{out}}(u+\epsilon)\partial_{u}\hat{\phi}_{\text{out}}(u)\ket{0}_{\text{out}}\right],
\end{align}
%%%
where we  subtracted the contribution of the zero-point fluctuations of the out-vacuum state.
From the relationship between the incident and reflected waves, $\hat{\phi}_{\text{out}} = \hat{\phi}_{\text{in}}\left(p(u)\right)$, we obtain
%%%
\begin{align}
    \bra{0}_{\text{in}}\partial_{u}\hat{\phi}_{\text{out}}(u_1)\partial_{u}\hat{\phi}_{\text{out}}(u_2)\ket{0}_{\text{in}} = -\frac{\partial_{u_1}p(u_1)\partial_{u_2}p(u_2)}{4\pi\left(p(u_1)-p(u_2)-i\epsilon\right)^2}.
\end{align}
%%%
Thus, we obtain the expectation value of the reflected wave as follows:
%%%
\begin{align}\label{eq: expectation value}
    \bra{0}_{\text{in}}\hat{T}_{uu}\ket{0}_{\text{in}} = -\frac{1}{24\pi}\left[\frac{\partial_{u}^3 p(u)}{\partial_u p(u)} - \frac{3}{2}\left(\frac{\partial_u^2p(u)}{\partial_up(u)}\right)^2\right].
\end{align}
%%%
%With the ray-tracing function  Eq.\eqref{eq: ray-tracing}, we can expect that Eq.\eqref{eq: expectation value} depends on Hawking temperature $T = \kappa/2\pi$. \anno{[?? please show expression of (6) for the ray-tracing functions (3)]} Practically\anno{[cannot understand meaning of this sentence]}, we can recognize that the expectation value becomes $\bra{0}_{\text{in}}\hat{T}_{uu}\ket{0}_{\text{in}} \sim \frac{\pi}{12} T^2$ in the region asymptotically $v=0$. Therefore, the reflected wave from the accelerating mirror corresponds to thermal radiation \anno{[why thermal?]} with temperature $T$.
Using the ray-tracing function \eqref{eq: ray-tracing}, Eq.~\eqref{eq: expectation value} becomes
%%%
\begin{align}\label{eq: g2}
\bra{0}_{\text{in}}\hat{T}_{uu}\ket{0}_{\text{in}} = \frac{\kappa^2}{48\pi}\frac{e^{\kappa u}\left(e^{\kappa u} + 2\right)}{\left(e^{\kappa u} + 1\right)^2} \approx \frac{\kappa^2}{48\pi} = \frac{\pi}{12}T_H^2\quad\text{for}\quad u\rightarrow +\infty,
\end{align}
%%%
where we  introduced the Hawking temperature $T_H: = \kappa/2\pi$. Equation \eqref{eq: g2} shows that an observer at future null infinity $\mathcal{I}^+ ~(v=\infty)$ sees the thermal flux with temperature $T_H$. Thus, we can identify the reflected wave mode as Hawking radiation.

%%%%%%%%%%%%%%%%%%%%%%%%%%%%%%%%%%%%%%%%%%%%%%%%%%%%%%%%%%%%%%%%
%%%%%%%%%%%%%%%%%%%%%%%%%%%%%%%%%%%%%%%%%%%%%%%%%%%%%%%%%%%%%%%%

\subsection{Entanglement-harvesting protocol with two-qubit detectors}\label{subsec: harvesting protocol}

In this section, we briefly introduce an entanglement-harvesting protocol using two-qubit detectors. This harvesting protocol was used to extract the entanglement from a quantum field \cite{Liu-Zhang-Yu-2023, Yoshimura-Tjoa-Mann-2021, Tjoa-Mann-2021, Cong-Tjoa-Mann-2019,Numbu-Ohsumi-2011,Nambu2013}. The description in this section is based on \cite{Numbu-Ohsumi-2011}.
We prepare two-qubit detectors, A and B, which interact with a massless scalar field (see the right panel of Fig.~\ref{fig: mirror model} for the setup). The detectors have two energy-level states, $\{\ket{\downarrow}, \ket{\uparrow}\}$, and their energy difference is denoted by $\Omega>0$.
The interaction Hamiltonian was assumed to take the following form: %\anno{[dervative coupling?]}
%%%
\begin{align}\label{eq: int. Hamiltonian}
  \hat{V}_{\text{int}} = \sum_{j = A,B}g(t)\left(e^{i\Omega t}\,\hat{\sigma}_j^{(+)} + e^{-i\Omega t}\,\hat{\sigma}_j^{(-)}\right)\hat\pi(t,x_j),
\end{align}
%%%
where $\hat\pi(t,x_j) = \partial_t\hat\phi(t,x_j)$ and $\hat{\sigma}^{\left(\pm\right)}$ are the ladder operators defined by
$ \hat{\sigma}^{(+)} = \ketbra{\uparrow}{\downarrow},~\hat{\sigma}^{(-)} = \ketbra{\downarrow}{\uparrow}$
with the switching function $g(t)$ of the interaction. In this study, we chose the Gaussian switching function
%%%
\begin{align}\label{eq: switching function}
  g(t) = \lambda \exp\left(-\frac{(t-t_0)^2}{2\sigma^2}\right),
\end{align}
%%%
where $\lambda$ is the coupling constant between the field and detectors. We prepared the initial down state $\ket{\downarrow}$ of the detectors and the initial vacuum state of the quantum field $\ket{0}$ at $t\to -\infty$ (initial time). The initial total state is $\ket{\Phi_0} =\ket{\downarrow}_A\ket{\downarrow}_B\ket{0}$ and evolves with the interaction \eqref{eq: int. Hamiltonian}. The final total state becomes 
%%%
\begin{align}\label{eq:qubit-state}
  \ket{\Phi} = \left[1 - i\int_{-\infty}^{\infty} dt_1\hat{V}_{\text{int}}(t_1) - \frac{1}{2}\int_{-\infty}^{\infty}dt_1dt_2\, T\left[\hat{V}_{\text{int}}(t_1)\hat{V}_{\text{int}}(t_2)\right] + O(g^3)\right]\ket{\Phi_0},
\end{align}
%%%
where $T\left[\cdots\right]$ denotes the time ordering. After the interaction, the detector density matrix is
%%%
\begin{align}\label{eq:density matrix}
  \rho_{AB} = \Tr_{\phi}\ketbra{\Phi} = 
  \begin{pmatrix}
    1 - E_A - E_B - X_4 & 0 & 0 & X \\ 
    0 & E_A & E_{AB} & 0 \\ 
    0 & E_{AB}^* & E_B & 0 \\ 
    X^* & 0 & 0 & X_4
  \end{pmatrix},
\end{align}
%%%
where we adopt the basis of this state $\left\{\ket{\downarrow_A\downarrow_B},\ket{\uparrow_A\downarrow_B},\ket{\downarrow_A\uparrow_B},\ket{\uparrow_A\uparrow_B}\right\}$.
The matrix components are as follows.
%%%
\begin{align}
  E_{j} &= \int^{\infty}_{-\infty}dt_1\int^{\infty}_{-\infty}dt_2 \,g(t_1)g(t_2)e^{-i\Omega (t_1-t_2)}\langle\hat{\pi}(t_1,x_j)\hat{\pi}(t_2,x_j)\rangle, \quad j = A,B \label{eq:dint1}\\ 
  E_{AB} &= \int^{\infty}_{-\infty}dt_1\int^{\infty}_{-\infty}dt_2 \,g(t_1)g(t_2)e^{-i\Omega (t_1-t_2)}\langle\hat{\pi}(t_1,x_A)\hat{\pi}(t_2,x_B)\rangle, \label{eq:dint2}\\ 
  X &= -2\int^{\infty}_{-\infty}dt_1\int^{t_1}_{-\infty}dt_2 \,g(t_1)g(t_2)e^{i\Omega (t_1+t_2)}\langle\hat{\pi}(t_1,x_A)\hat{\pi}(t_2,x_B)\rangle, \label{eq:dint3}\\ 
  X_4 &= 4\int^{\infty}_{-\infty}dt_3\int^{t_3}_{-\infty}dt_4\int^{\infty}_{-\infty}dt_1\int^{t_1}_{-\infty}dt_2 \,g(t_3)g(t_4)g(t_1)g(t_2)e^{i\Omega (t_1+t_2)} \notag \\ 
  &\hspace{4cm}\times \langle\hat{\pi}(t_4,x_B)\hat{\pi}(t_3,x_A)\hat{\pi}(t_1,x_A)\hat{\pi}(t_2,x_B)\rangle.\label{eq:dint4}
\end{align}
%%%
Positivity of the state $\rho_{AB}$ requires the following conditions for the components:
%%%
\begin{equation}
 E_AE_B\ge|E_{AB}|^2,\quad X_4\ge |X|^2.
 \label{eq:positivity}
\end{equation}
$E_A$ and $E_B$ represent the excitation probabilities of detectors A and B, respectively, and $X_4$ represents the probability of simultaneous excitation of A and B. In particular, as we will see, $X_4$ is related to second-order coherence \eqref{eq:g2}, hence, it is a key quantity in this study. $E_{AB}$  is related to the first-order coherence. $X$ represents the coherence between states $\ket{\downarrow_A\downarrow_B}$ and $\ket{\uparrow_A\uparrow_B}$ and represents the non-local correlation between A and B related to entanglement. The diagonal components of $\rho_{AB}$ can be obtained directly as the response (output) of the detectors, whereas the off-diagonal components cannot. Quantum tomography of the two-detector state must be performed to obtain these off-diagonal components \cite{D'Ariano-Paris-Sacchi-2003}.

Using Wick's theorem \cite{Wick-1950}, the four-point function in $X_4$ can be described by the following two-point functions:
%%%
\begin{align*}
  \langle\hat{\pi}(t_4,x_B)\hat{\pi}(t_3,x_A)\hat{\pi} (t_1,x_A)\hat{\pi}(t_2,x_B)\rangle &= \langle\hat{\pi}(t_4,x_B)\hat{\pi}(t_1,x_A)\rangle\langle\hat{\pi}(t_3,x_A)\hat{\pi}(t_2,x_B)\rangle \\ &\qquad+\langle\hat{\pi}(t_4,x_B)\hat{\pi}(t_3,x_A)\rangle\langle\hat{\pi}(t_1,x_A)\hat{\pi}(t_2,x_B)\rangle\\
  &\qquad+\langle\hat{\pi}(t_3,x_A)\hat{\pi}(t_1,x_A)\rangle\langle\hat{\pi}(t_4,x_B)\hat{\pi}(t_2,x_B)\rangle. 
\end{align*}
%%%
From this relationship, we can rewrite $X_4$ using $E_{A,B}, E_{AB}$ and $X$ as follows:
%%%
\begin{align}\label{eq: X4}
  X_4 = E_AE_B + |E_{AB}|^2 + \abs{X}^2.
\end{align}
%%%
Thus, one of the positivity condition \eqref{eq:positivity} of the state is automatically satisfied.
We change the integration variables $t_1, t_2$  to $x, y$ as
$x = (t_1 + t_2)/2, y = (t_1 - t_2)/2$.
Then, the components of the two detector state are expressed as
%%%
\begin{align*}
  E_{AB}  &=  2\lambda^2 \int^{\infty}_{-\infty}dx\int^{\infty}_{-\infty}dy\, e^{-\frac{x^2}{\sigma^2}} e^{-\frac{y^2}{\sigma^2}-2i\Omega y}D(x+t_0,y,x_A,x_B), \\ 
  %E_{j}   &=  E_{AB}(d=0), \anno{[??]}\\
  X       &=  -4\lambda^2 \int_{-\infty}^{\infty}dx \int_{0}^{\infty}dy\, e^{-\frac{x^2}{\sigma^2}+2 i \Omega x} e^{-\frac{y^2}{\sigma^2}} D(x+t_0,y,x_A,x_B), 
\end{align*}
%%%
where $D(x+t_0,y,x_A,x_B)$ is the Wightman function for the derivative of the scalar field. %\anno{[dervative coupling?]}
%%%
\begin{align}
  D(x+t_0,y,x_A,x_B): = \langle\hat{\pi}(t_1,x_A)\hat{\pi}(t_2,x_B)\rangle,\quad x+t_0=\frac{t_1+t_2}{2},\quad y=\frac{t_1-t_2}{2}.
\end{align}
%%%
The excitation probabilities of each detector were calculated as $E_A=E_{AB}(x_B\rightarrow x_A)$ and $E_B=E_{AB}(x_A\rightarrow x_B)$.
%With the mirror trajectory given by Eq.\eqref{eq: ray-tracing},\anno{[?? the following form is not specific to the trajectory (3)]}  the Wightman function is give by \cite{Birrell-Davis-1984}
In the moving-mirror model, the Wightman function is expressed using the ray-tracing function \cite{Birrell-Davis-1984}:%\anno{[dervative coupling?]}
%%%
\begin{align}
  D(x,y,x_A,x_A+d) &= -\frac{1}{4\pi}\left[\frac{p'(u_A)p'(u_B)}{\left(p(u_A) - p(u_B) - i\epsilon\right)^2} + \frac{1}{\left(v_A-v_B-i\epsilon\right)^2} \right. \notag \\ 
  &\qquad\qquad\left.- \frac{p'(u_A)}{\left(p(u_A) - v_B - i\epsilon\right)^2} - \frac{p'(u_B)}{\left(v_A - p(u_B) - i\epsilon\right)^2}\right].
  \label{eq:wightman}
\end{align}
%%%
where $u_A=t_1-x_A, v_A=t_1+x_A, u_B=t_2-(x_A+d), v_B=t_2+(x_A+d), d=x_B-x_A$ and $\epsilon > 0$ is the UV cut-off parameter. 
For the ray-tracing function, Eq.~\eqref{eq: ray-tracing}, the first term of the Wightman function ~\eqref{eq:wightman} represents Hawking radiation at future null infinity $\mathcal{I}^+$ ($v_{A,B}\rightarrow +\infty$):
%%%
\begin{equation}
D_1=
\begin{cases}
-\dfrac{1}{4\pi}\dfrac{1}{(u_A-u_B-i\ep)^2},&\quad u\rightarrow -\infty \\
-\dfrac{\kappa^2}{16\pi}\dfrac{1}{\left[\sinh(\kappa/2)(u_A-u_B)-i\ep\right]^2},&\quad u\rightarrow+\infty
\end{cases}.
\label{eq:D1}
\end{equation}
%%%
For $u\rightarrow+\infty$, contribution to the excitation probability $E_A$ is calculated as  
%%%
\begin{align}
    E_A(\Omega)=\frac{\lambda^2}{2\pi}\int_{-\infty}^{+\infty} dk \frac{e^{-(k-\Omega\sigma)^2}k}{e^{2\pi k/(\kappa \sigma)}-1} \approx\frac{\lambda^2}{2\sqrt{\pi}} \frac{\Omega\sigma}{e^{2\pi\Omega/\kappa}-1}\quad \text{for}\quad \kappa\sigma\gg 1.%\quad \anno{\text{[OK?]}}
    \label{eq:Planck}
\end{align}
%%%
It shows a Planckian distribution with the Hawking temperature $T_H=\kappa/2\pi$; hence, we can recognize the thermal radiation from the accelerating mirror.

Using the density matrix \eqref{eq:density matrix}, we can evaluate the quantities related to the quantum information of the system. A typical example is negativity $\mathcal{N}$ \cite{Vidal-Werner-2002} that determines the existence of bipartite entanglement. This is defined by the partial transposition of the detector density matrix. In our setup, the negativity can be expressed as 
%%%
\begin{align}\label{eq:negativity}
  \mathcal{N} = \mathrm{max}\left[\sqrt{\abs{X}^2 + \left(\frac{E_A-E_B}{2}\right)^2} - \frac{E_A+E_B}{2},0\right]\ge 0.
\end{align}
%%%
If the negativity $\mathcal{N}$ is non-zero, a quantum entanglement exists between the two detectors. The negativity represents amount of entanglement; in fact, logarithmic negativity $\ln(2\mathcal{N}+1)$ provides an upper bound of distillable entanglement (the number of the Bell pairs extractable from a bipartite state) \cite{Vidal-Werner-2002}. For a two-qubit system, $\mathcal{N}>0$ provides the necessary and sufficient condition for the entanglement (non-separability) of the bipartite state. The method of estimating the entanglement of the quantum field using the detector state is called the entanglement-harvesting protocol \cite{Simidzija-Martinez-2017,Simidzija-Martin-Martinez-2018}. %\anno{[refs]}
The positivity of the negativity \eqref{eq:negativity} is equivalent to
%%%
\begin{align}\label{eq:N>0}
    \mathcal{N} > 0\quad \Leftrightarrow\quad \abs{X}^2 > E_A E_B.
\end{align}
%%%
Thus, in this study, we introduce the following quantity to judge the existence of entanglement:
%%%
\begin{equation}
\widetilde{\mathcal{N}}:=\frac{|X|^2}{E_AE_B}-1.
\label{eq:tneg}
\end{equation}
%%%
This quantity is independent of the strength of the coupling constant $\lambda$ and is suitable for the numerical evaluation of  entanglement.
For a fixed value of $|X|$, as the product $E_AE_B$ increases, the negativity decreases and the quantum correlation (entanglement) decreases and  vanishes at a certain value of $E_AE_B$. As $E_{A}$ and $E_{B}$ represent the transition probability of each detector due to vacuum fluctuation or thermal Hawking radiation, larger $E_AE_B$ corresponds to larger local ``noise" which will destroy quantum coherence  and quantum correlations between detectors.

%%%%%%%%%%%%%%%%%%%%%%%%%%%%%%%%%%%%%%%%%%%%%%%%%%%%%%%%%%%%%%%%
%%%%%%%%%%%%%%%%%%%%%%%%%%%%%%%%%%%%%%%%%%%%%%%%%%%%%%%%%%%%%%%%
%%%%%%%%%%%%%%%%%%%%%%%%%%%%%%%%%%%%%%%%%%%%%%%%%%%%%%%%%%%%%%%%

\section{Coherence of scalar field}\label{sec. coherence}

\subsection{First-order coherence and second-order coherence}\label{subsec: g2}
\begin{figure}[t]
    \centering
    \includegraphics[width=0.8\linewidth]{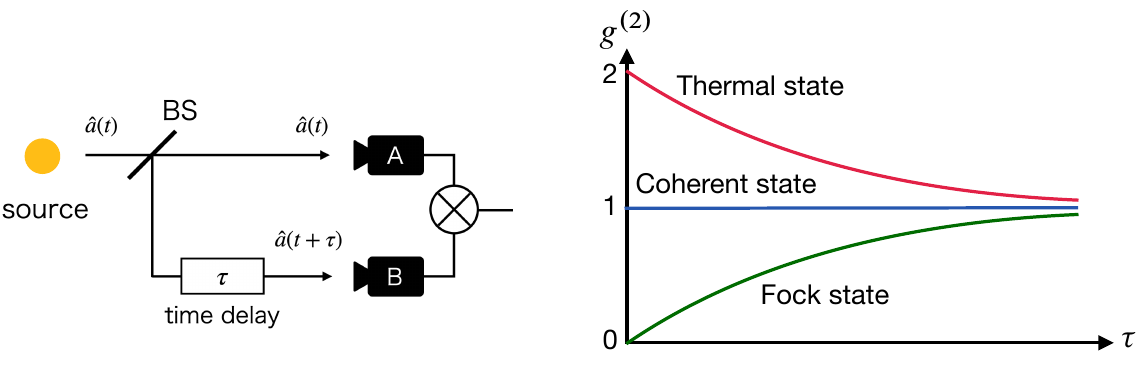}
    \caption{Left panel: HBT experiment for a single-mode photon.  The photon emitted from  a light source is then divided by a beam splitter (BS). One photon enters the detector A, and another enters the detector B with a delay time $\tau$. The correlation of photon numbers between A and B is measured and the intensity correlation $g^{(2)}$ (second-order coherence) is obtained. Right panel: $\tau$ dependence of $g^{(2)}$. The plot shows the nature of the quantum state of the light source as a different behavior of $g^{(2)}$. The limiting value $g^{(2)}(0)$ discriminates the states.}
    \label{fig:g2}
\end{figure}

\begin{figure}[t]
  \centering
  \includegraphics[width=0.48\linewidth]{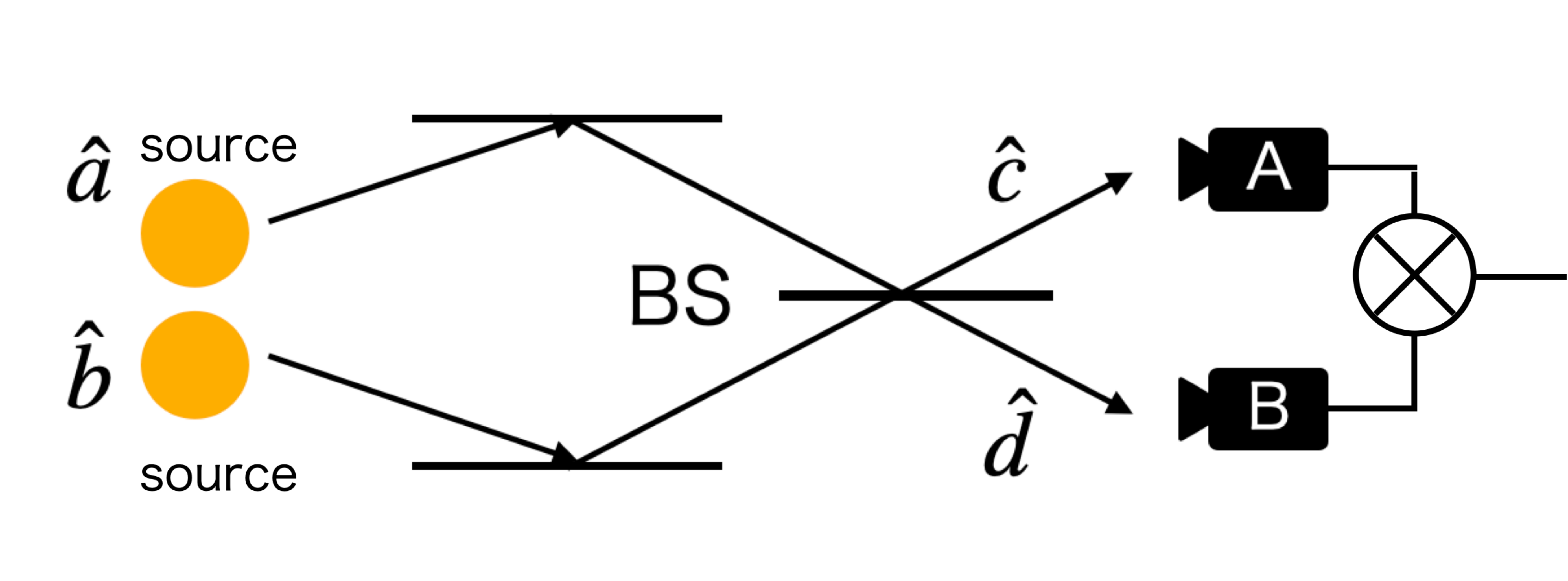}
  \caption{HBT experiment using a light source with two modes $\hat a$ and $\hat b$. They enter the BS and produce the output modes $\hat{c}$ and $\hat{d}$. The output modes can be entangled by the action of the BS.}
  \label{fig: HBT two}
\end{figure}

Higher-order coherence of light is used to clarify the quantum nature of light sources in quantum optics. In interferometry experiments with a light source, Young's double slit experiment \cite{Carnal-Mlynek-1991} and Mach-Zehnder interferometry \cite{Paris-1999} characterize the first-order coherence of light $g^{(1)}$ defined by
%%%
\begin{equation}
g^{(1)}(\tau):=\frac{\expval{\hat a^\dag(t+\tau)\hat a(t)}}{\sqrt{\expval{\hat a^\dag(t)\hat a(t)}}\sqrt{\expval{\hat a^\dag(t+\tau)\hat a(t+\tau)}}},
\end{equation}
%%%
where $\hat a(t)$ denotes the annihilation operator for a single light mode and $\tau$ denotes a path difference or a time delay. These experimental setups are useful for measuring the classical interference of light. However, they cannot reveal the quantum nature of light sources encoded in the higher-order coherence of  light. The Hanbury-Brown and Twiss (HBT) effect \cite{Hanbury-Brown-Twiss-1954, Fano-1961} reveals the quantum nature of light. This effect was originally introduced in the field of astronomical observation. HBT measured the intensity correlation of light from a star using two detectors (antennas) and estimated the apparent diameter of the star. They also conducted an experiment to clarify the physical meaning of the intensity correlation of light. 
%%%
Figure \ref{fig:g2} shows the experimental setup used to understanding the physical meaning of the intensity correlation of  single-mode photons. 
As shown in the right-hand panel of Fig.~\ref{fig:g2}, the quantum state of light source can be discriminated by measuring the intensity correlation of light.  In this setup with a single-mode light source, the second-order coherence $g^{(2)}$ which quantifies the intensity correlation, is defined as
%%%
\begin{equation}
g^{(2)}(\tau):=\frac{\expval{\hat a^\dag(t)\hat a^\dag(t+\tau)\hat a(t+\tau)\hat a(t)}}{\expval{\hat a^\dag(t)\hat a(t)}\expval{\hat a^\dag(t+\tau)\hat a(t+\tau)}},
\end{equation}
%%%
where $\tau$ denotes a time delay.
As shown in the right-hand panel of Fig.~\ref{fig:g2}, $g^{(2)}(0)$ takes different values depending on the quantum state of the light source, and it is possible to determine  the quantumness of light. For $1\le g^{(2)}(0)\le 3$, a normalizable P-function which is the  classical counter-part to the quantum distribution function \cite{Gardiner2004} exists. If the P-function exists, we can say that the state is ``classical''. In fact, the coherent state ($g^{(2)}(0)=1$) and  thermal state ($g^{(2)}(0)=2$) are P-representable, and they are ``classical'' states. The states with $2<g^{(2)}(0)$ are squeezed lights, whereas the state with $3<g^{(2)}(0)$ is not P-representable. 
%%%
%\begin{figure}[ht]
%%%
 Now, let us consider a light source with two mode, $\hat a$ and $\hat b$ (see Fig.~\ref{fig: HBT two}). 
 They enter a BS and produce  output modes $\hat{c}$ and $\hat{d}$. The unitary operation of the BS can be expressed as $\hat U_{BS} = e^{i\theta \left(\hat{a}^{\dagger}\hat{b} + \hat{a}\hat{b}^{\dagger}\right)}$. Therefore, the output modes are represented by $\hat{c} = r\hat{a} + t\hat{b}, \hat{d} = -t^{*}\hat{a} + r^{*}\hat{b}$, where $r$ and $t$ are the reflection and transmission coefficients, respectively. If we prepare a 50:50 BS, the coefficients are $r = t = 1/\sqrt{2}$. Thus, we obtain the following modes as the output of the BS:
%%%
\begin{align}
    \hat{c} = \frac{1}{\sqrt{2}}\left(\hat{a}+\hat{b}\right),\quad \hat{d}= \frac{1}{\sqrt{2}}\left(-\hat{a}+\hat{b}\right).
\end{align}
%%%
%%%
%%%
\noindent
For a non-classical state of the input modes, the output modes can be entangled. In \cite{Strobinska-Wodkiewicz-2005,Cotler-Wilczek-Borish-2021}, it was noted that the entanglement between the output modes $\hat{c}$ and $\hat{d}$ could be evaluated using the second-order coherence $g^{(2)}$ defined by
%%%
\begin{align}
  g^{(2)} = \frac{\langle\hat{I}_A\hat{I}_B\rangle}{\langle \hat{I}_A\rangle\langle \hat{I}_B\rangle},\quad
  \hat I_A=\hat c^\dag\hat c,\quad
  \hat I_B=\hat d^\dag\hat d,
  \label{eq:g2}
\end{align}
%%%
where $\langle \hat{I}_j\rangle,j=A,B$ is the intensity measured by detectors A (or B), and $\langle \hat{I}_A\hat{I}_B \rangle$ is the intensity correlation between $\hat I_A$ and $\hat I_B$. 

For the two-qubit detector system, the intensity can be obtained directly as the excitation probability of the qubit detectors \eqref{eq:qubit-state}. Therefore, we rewrite $\langle \hat{I}_A \rangle = E_A$, $\langle \hat{I}_B \rangle = E_B$, and $\langle\hat{I}_A\hat{I}_B \rangle = X_4$.
From this, Eq.~\eqref{eq:g2} can be rewritten as  follows:
%%%
\begin{align}\label{eq:g22}
  g^{(2)} =\frac{X_4}{E_AE_B}= 1 + \frac{|E_{AB}|^2}{E_AE_B}+ \frac{\abs{X}^2}{E_AE_B},
\end{align}
%%%
where we used Eq.~\eqref{eq: X4}. 
This is the key equation used in our analysis of the quantum nature of Hawking radiation. 
In this formula, the first-order coherence is also included as
%%%
\begin{equation}
 g^{(1)}=\frac{E_{AB}}{\sqrt{E_AE_B}}.
\end{equation}
%

%%%%%%%%%%%%%%%%%%%%%%%%%%%%%%%%%%%%%%%%%%%%%%%%%%%%%%%%%%%%%%%%
%%%%%%%%%%%%%%%%%%%%%%%%%%%%%%%%%%%%%%%%%%%%%%%%%%%%%%%%%%%%%%%%

\subsection{Comparison with negativity}\label{subsec: N vs g2}
\begin{figure}[t]
  \centering
  \includegraphics[width=0.28\linewidth]{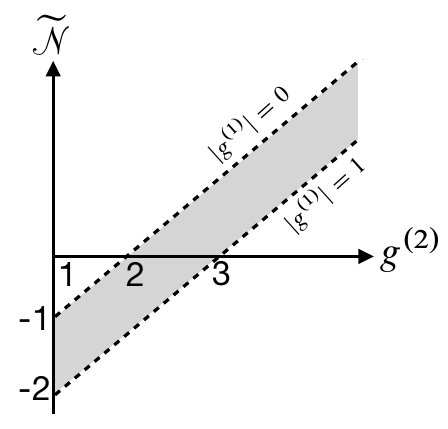}
  \caption{
  The schematic diagram of the relation between values of $g^{(2)}$ and $\widetilde{\mathcal{N}}$. In the $(g^{(2)},\widetilde{\mathcal{N}})$-plane, the shaded region corresponds to an allowed parameter range. For $1<g^{(2)}<2$,  $\widetilde{\mathcal{N}}<0$ and AB is separable. For $3<g^{(2)}$, $\widetilde{\mathcal{N}}>0$ and AB is entangled and entanglement-harvesting succeeds. For $2<g^{(2)}<3$, we cannot determine the sign of the negativity without a value of the first-order coherence $g^{(1)}$. }
  \label{fig: g2 value}
\end{figure}

%%%
 In this section, we  compare the behavior of  second-order coherence with that of negativity.
Using the expression for $g^{(2)}$ \eqref{eq:g22}, and the entanglement condition \eqref{eq:N>0}, we can relate the values of these two quantities. That is, a combination of degrees of coherence is equal to negativity:
%%%
\begin{equation}
 g^{(2)}-\left|g^{(1)}\right|^2=\widetilde{\mathcal{N}}+2,
\end{equation}
%%%
and existence of entanglement of the two-qubit state is equivalent to the following inequality for degrees of coherence:
%%%
\begin{equation}
 \widetilde{\mathcal{N}}>0\quad\Leftrightarrow\quad g^{(2)}-\left|g^{(1)}\right|^2>2.
\end{equation}
%%%
Note that the first-order coherence $g^{(1)}$ takes values smaller than unity because $E_{AB}$ is defined as a two-point function of detectors A and B, whereas  $E_{A}$  and $E_B$ are defined as auto-correlation functions (local noise). This is equivalent to the one of the positivity condition of the two-qubit state \eqref{eq:positivity}.
Thus,  even if we do not have knowledge about a value of $g^{(1)}$, we obtain the following range for $g^{(2)}$: 
%%%
\begin{align}\label{condition: g2 with EA>EAB}
  2 + \widetilde{\mathcal{N}} < g^{(2)} < 3 + \widetilde{\mathcal{N}}.
\end{align}
%%%
The lower and upper bounds of $g^{(2)}$ depend on the values of $\abs{X}$, $E_A$, $E_B$. 
In addition,  we have the following relationship between the values of $g^{(2)}$ and $\widetilde{\mathcal{N}}$: 
%%%
\begin{align}
  1 < g^{(2)} < 2 \quad &\Rightarrow \quad \widetilde{\mathcal{N}} < 0, \label{eq:tnless0}\\ 
  3<g^{(2)} \quad &\Rightarrow \quad  \widetilde{\mathcal{N}} > 0.
  \label{eq:tngtr0}
\end{align}
%%%
This relationship is illustrated in Fig.~\ref{fig: g2 value}.
For $1<g^{(2)}<2$, we can conclude that $\widetilde{\mathcal{N}}<0$ and AB are separable.  On the other hand, for $3<g^{(2)}$, it is possible to conclude that $\widetilde{\mathcal{N}}>0$ and AB are entangled. Thus, we can estimate the quantum entanglement of the bipartite system AB of two-qubit detectors using  the second-order coherence.  For $2<g^{(2)}<3$, we cannot say anything about the entanglement from the value of $g^{(2)}$ without knowing of $g^{(1)}$.
%%%
%%%

 In our setup of the moving mirror model, $3<g^{(2)}$  corresponds to the situation in which Hawking radiation is represented as a squeezed thermal state and is not P-representable. In this sense, the state is ``quantum''.
On the other hand, the state with a value of $2<g^{(2)}<3$ is ``classical''  because the state is P-representable \cite{Strobinska-Wodkiewicz-2005}. Although we cannot judge the separability simply by examining  the value of $g^{(2)}$, if the first-order coherence  satisfies $|g^{(1)}|\ll 1$, then, we have the following approximate relation between the intensity correlation and the negativity:
%%%
\begin{equation}
 g^{(2)}\approx \widetilde{\mathcal{N}}+2.
 \label{eq:g2tN}
\end{equation}
%%%
Therefore, the second-order coherence is a good indicator of entanglement of the two-qubit state in this case.
%%%

%%%%%%%%%%%%%%%%%%%%%%%%%%%%%%%%%%%%%%%%%%%%%%%%%%%%%%%%%%%%%%%%
%%%%%%%%%%%%%%%%%%%%%%%%%%%%%%%%%%%%%%%%%%%%%%%%%%%%%%%%%%%%%%%%
\newpage
\section{Numerical Results and discussions}\label{sec. result}
This section presents the numerical results of entanglement negativity and degrees of coherence for the two-qubit state.

\subsection{Saddle point approximation of the state}
\begin{figure}[t]
\centering
\includegraphics[width=0.8\linewidth]{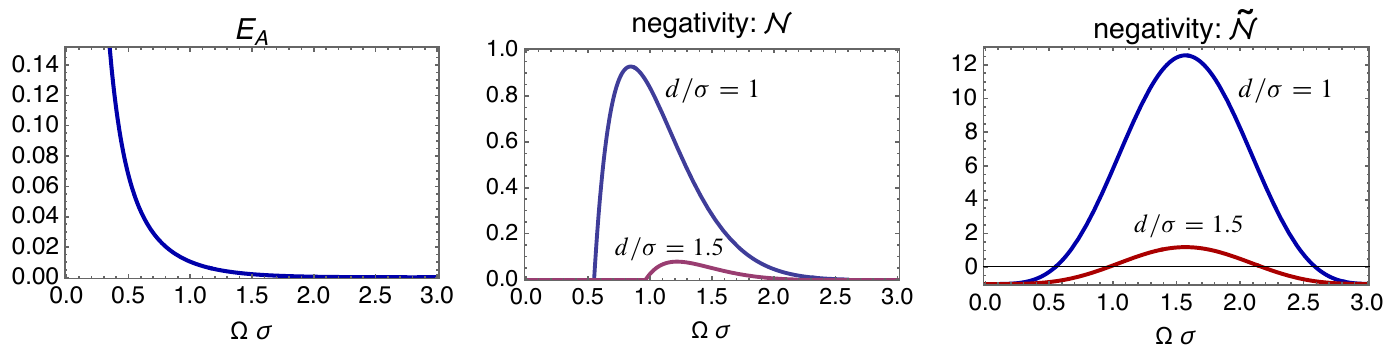}
\caption{Local noise $E_A$ and negativity $\mathcal{N},\widetilde{\mathcal{N}}$ for the Wightman function $D_1$ (Eq.~\eqref{eq:D1}), which represents Hawking radiation measured at $\mathcal{I}^+$. The parameters are $\kappa\sigma=1, \lambda^2=4$. This choice of $\lambda^2$ is equivalent to ploting re-scaled values of $E_A$ and $\mathcal{N}$: $4E_A/\lambda^2, 4\mathcal{N}/\lambda^2$. Both $\mathcal{N}$ and $\widetilde{\mathcal{N}}$  provide the same separability condition  of the state, which is independent of $\lambda^2$. }
\label{fig:EAneg}
\end{figure}

It is technically difficult to evaluate the components of the two-detector state using Eq.~\eqref{eq:density matrix} because the quantities are expressed as double integrals of the oscillatory functions with respect to the time variable. 
Thus, in this study, we analyze $g^{(2)}$ and $\mathcal{N}$ using numerical calculations with a saddle-point approximation. Saddle-point approximation is a method for obtaining the approximate values of integrals in a complex plane. Consider the following line integral along the path $C$:
%%%
\begin{align}
    I = \int_C dz\, f(z)e^{g(z)},
\end{align}
%%%
where $f(z)$ and $g(z)$ are the analytic functions in the complex plane. The path must be chosen to be closed, or the real part of $g(z)$ will be negatively infinite at both endpoints of the path, and the integration function will become zero. 
If the real part of $g(z)$ takes its maximum value at the point $z_0$ in the complex plane, we call this point the saddle point.
By evaluating the Gaussian integral for $z_0$, the integral can be approximated as 
%%%
\begin{align}
    I \approx \frac{\sqrt{2\pi}f(z_0)e^{g(z_0)}e^{i\alpha}}{\sqrt{\abs{f''(z_0)}}}.
\end{align}
%%%%
In the saddle-point approximation, the integral is performed near the saddle point $z_0=x_0+iy_0$. However, we must be careful about the applicability of this method to our problem because the Wightman function has an infinite number of singular points (poles) in the upper half of the complex plane. There are several methods that have been introduced to provide numerical calculations for such functions \cite{Salton-Mann-Menicucci-2015, Tjoa-Mann-2021}, but we will use the method \cite{Numbu-Ohsumi-2011,Nambu2013, Salton-Mann-Menicucci-2015} that includes the contribution of one pole in the integral. By applying the saddle-point approximation to our setup, we obtain the components of
Eq.~\eqref{eq:density matrix}:
%%%
\begin{align}
  E_{AB} &=  2 \pi \lambda^2 \sigma^2 e^{-\left(\Omega \sigma\right)^2}D(t_0, -i\Omega \sigma^2, x_A,x_A+d), \\ 
  X      &= -2 \pi \lambda^2 \sigma^2 e^{-\left(\Omega \sigma\right)^2}D(t_0 + i\Omega \sigma^2, 0, x_A,x_A+d),
\end{align}
%%%
where $D(x,y,x_A,x_A+d)$ is defined by Eq.~\eqref{eq:wightman}:
The first term in $D(x,y,d)$  has the largest contribution, and for the ray-tracing function \eqref{eq: ray-tracing}; it contains an infinite number of poles in the upper half of a complex plane. After applying the saddle-point approximation, as shown in Eq.~\eqref{eq:wightman}, we have shifted the Wightman function by $\Omega \sigma^2$ to the direction of the imaginary axis.
If the shift is larger than the separation of the poles, our approximation becomes poor because we only consider the contribution of a single pole.
Thus, we should consider the condition in which the shift is smaller than the pole separation.
Because the separation of the poles is $\pi/\kappa$ in the first term of Eq.~\eqref{eq:wightman}, the following condition must be satisfied in our calculation (see \cite{Numbu-Ohsumi-2011,Nambu2013,Salton-Mann-Menicucci-2015} for a detailed calculation):
%%%
\begin{equation}
\kappa\, \Omega\, \sigma^2 < \pi.
\end{equation}
%%%
For the Wightman fucntion \eqref{eq:D1}, which is the main contribution to the components of the state on the future null infinity $\mathcal{I}^+$, 
%%%
\begin{align}
 E_{AB}&= (2\pi\lambda^2\sigma^2e^{-(\Omega\sigma)^2})\times
 \left(-\frac{\kappa^2}{16\pi\sinh^2\left[(\kappa/2)(d-2i\Omega\sigma^2)\right]}\right),\\
 |X|&=(2\pi\lambda^2\sigma^2e^{-(\Omega\sigma)^2})\times
 \left(\frac{\kappa^2}{16\pi\sinh^2\left[(\kappa/2)d\right]}\right),
\end{align}
%%%%
and negativity \eqref{eq:tneg} is given by
%%%
$
\widetilde{\mathcal{N}}=\left(\sin(\kappa\Omega\sigma^2)/\sinh(\kappa d/2)\right)^4-1$.
Degrees of coherence are
%%%
\begin{align}
g^{(1)}=-\left(\frac{\sin(\kappa\Omega\sigma^2)}{\sinh(\kappa(d/2-i \Omega\sigma^2))}\right)^2, \quad
g^{(2)}=1+\left|\frac{\sin(\kappa\Omega\sigma^2)}{\sinh(\kappa(d/2-i\Omega\sigma^2))}\right|^4+\left(\frac{\sin(\kappa\Omega\sigma^2)}{\sinh(\kappa d/2)}\right)^4.
\end{align}
Figure~\ref{fig:EAneg} shows the $\Omega$-dependence of local noise $E_A$ and negativity. The behavior of $E_A$ is consistent with the exact formula \eqref{eq:Planck}. Negativity becomes maximum at $\Omega=\pi/(2\kappa\sigma^2)$. The condition of entanglement is $\sinh(\kappa d/2)<\sin(\kappa\Omega\sigma^2)$, which implies that a large separation of two detectors destroys the entanglement. For a fixed value of $d$, negativity $\widetilde{\mathcal{N}}$ becomes negative for small values of $\Omega$, which corresponds to ``no-go theorem" of entanglement-harvesting with qubit detectors \cite{Simidzija-Martin-Martinez-2018}. In contrast, for large values of $\Omega$, negativity also becomes zero beyond the critical value of $\Omega$. Around this point, the negativity $\mathcal{N}$ decays exponentially as $e^{-\Omega^2\sigma^2}$ and becomes zero at the critical value of $\Omega$. Thus, the harvestable amount of entanglement is very small for a large $\Omega$ region.

%%%

\begin{figure}[H]
\centering
\includegraphics[width=0.52\linewidth]{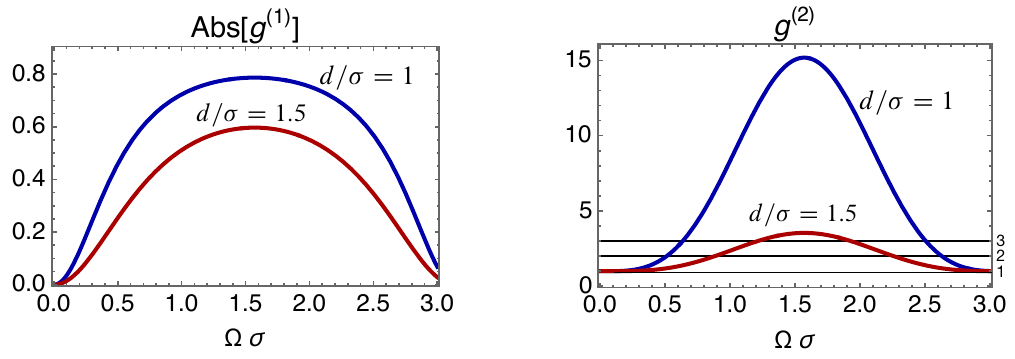}
\caption{$\Omega$-dependence of $g^{(1)}$ and  $g^{(2)}$ for the Wightman function $D_1$  (Eq.~\eqref{eq:D1}). The parameter is $\kappa \sigma=1$.}
\label{fig:g1g2D1}
\end{figure}

Figure \ref{fig:g1g2D1} shows the degrees of coherence as a function of $\Omega$. These quantities are related to the behavior of entanglement. The first-order coherence $g^{(1)}$ represents ``coherence" of two detectors; for a larger separation $d$, $g^{(1)}$ becomes smaller. At $\Omega=0$ and $\Omega=\pi/(\kappa\sigma^2)$, $g^{(1)}$ becomes zero due to the effect of the local noise $E_{A,B}$ (decoherence due to local noise). The second-order coherence $g^{(2)}$ also has a small value one about $\Omega=0,\pi/(\kappa\sigma^2)$, which behavior is correlated with that of $g^{(1)}$. As $|g^{(1)}|^2\ll1$ in this case, negativity and $g^{(2)}$ are related as $g^{(2)}\approx\tilde N+2$, and $g^{(2)}$ is a good indicator of entanglement.
%%%

\newpage
%%%%%%%%%%%%%%%%%%%%%%%%%%%%%%%%%%%%%%%%%%%%%%%%%%%
\subsection{Entanglement-harvesting}
\begin{figure}[h]
    \centering
       \includegraphics[width=0.7\linewidth]{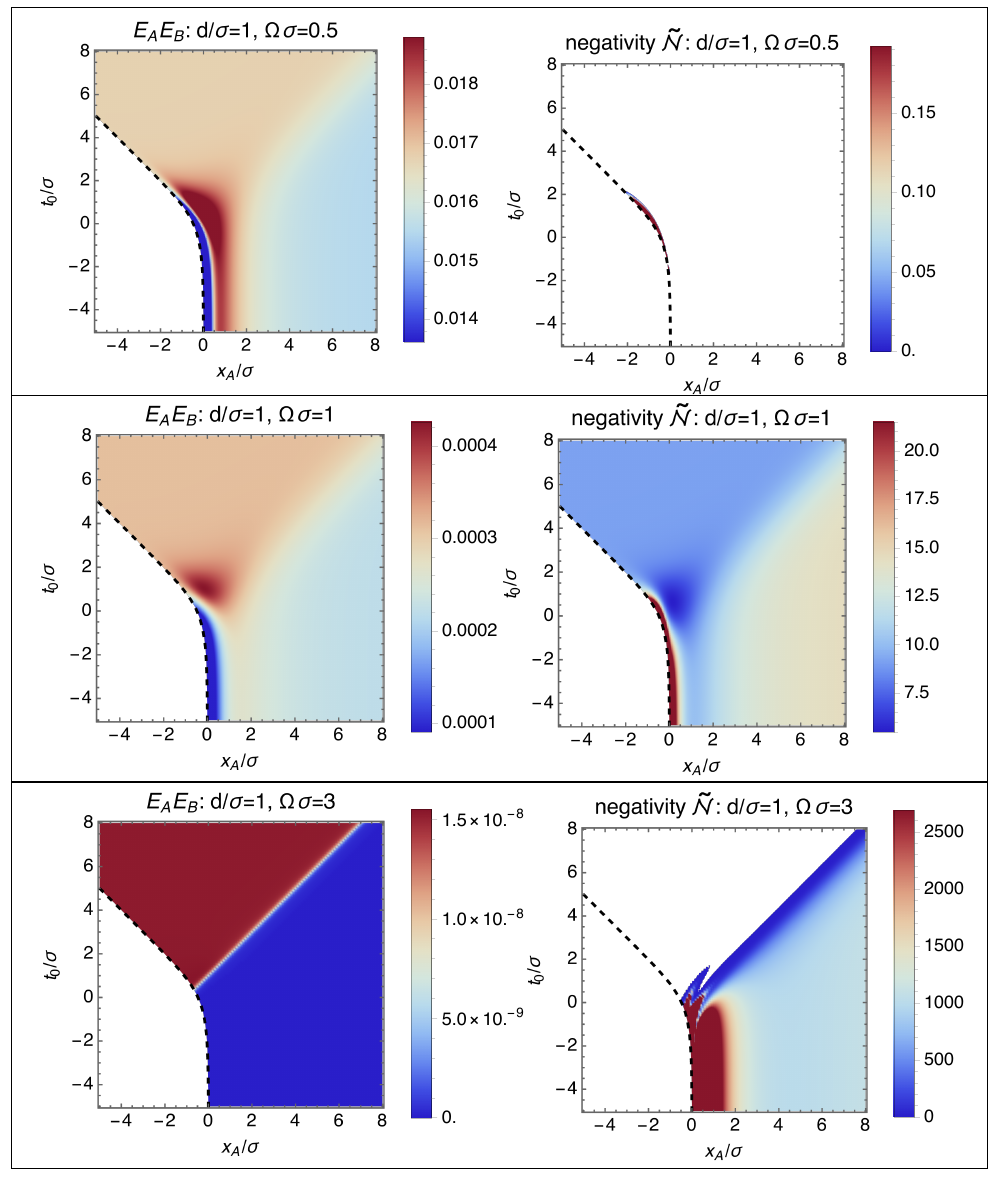} 
    \caption{ The distribution of $E_AE_B$ and $\widetilde{\mathcal{N}}$ in the $(t_0,x_A)$-plane with  $\Omega\sigma=0.5,1,3$ (from the top panels to the bottom panels). Other parameters are $d/\sigma=1, \kappa/\sigma=1,\lambda^2=4$. The dashed line represents a trajectory of the mirror. Regions with large values of $E_AE_B$ ($t_0/\sigma \sim1, x_A\sim 0$) reflect the emission of Hawking radiation from the mirror (a source region of Hawking radiation). The emitted radiation propagates with the speed of light towards  future null infinity $\mathcal{I}^+$. Hawking radiation reduces the amount of entanglement and for $\Omega\sigma=3$ case, it results in entanglement death.}
    \label{fig:noise-neg}
\end{figure}

Now we take into account all the terms in the Wightman function \eqref{eq:wightman} and evaluate the negativity of the two-qubit state in the moving mirror system.
Figure~\ref{fig:noise-neg} shows the distribution of noise $E_A E_A$ and negativity $\widetilde{\mathcal{N}}$ in the $(t_0,x_A)$-plane  with the saddle-point approximation of the two-qubit state. 
%%%
%\begin{figure}[ht]
%%%

The emission of Hawking radiation starts when the mirror begins accelerating, and this emission is reflected as large values of $E_AE_B$ (red-colored regions in the figures). The size of the red-colored regions in the figures increases with the speed of light. As Hawking radiation has a thermal spectrum, it may reduce or destroy the quantum coherence between detectors. Before the emission of Hawking radiation starts, we also have non-zero small values of $E_AE_B$, which reflects the existence of ``noise'' nothing to do with Hawking radiation (local vacuum fluctuation in Minkowski spacetime).  This noise also reduces or destroys quantum correlations.

Local fluctuations, such as vacuum fluctuations and Hawking radiation, act as noise that destroys quantum entanglement. Hence, with the change of the system parameters, the harvestable (detectable) entanglement will suddenly become zero due to an increase in local ``noise'', and this behavior is called entanglement death. According to Eq.~\eqref{eq:N>0}, the condition for entanglement death is $\abs{X}^2 < E_AE_B$, where $E_AE_B$ quantifies the amount of ``noise'' that breaks quantum entanglement. Several studies have investigated this phenomenon. For example, \cite{Cong-Tjoa-Mann-2019} pointed out that the entanglement could not be harvested near the mirror.  As shown in the right-hand panels of Fig.~\ref{fig:noise-neg}, entanglement harvesting is possible near the mirror in our setup. This is due to the differences in observation quantity adopted.  In previous research \cite{Cong-Tjoa-Mann-2019}, $\langle \hat\phi\, \hat\phi \rangle$ was adopted as the quantity to be measured by the detectors, while we assumed that $\langle\hat\partial_t\phi\,\partial_t\hat\phi\rangle$ is an observable quantity in this study. We impose the boundary condition $\hat\phi(t,x_{\text{mirror}}) = 0$ on the mirror and $\langle\hat\phi\,\hat\phi\rangle = 0$  but $\langle\partial_t\hat\phi\,\partial_t\hat\phi\rangle\neq 0$ on the mirror. 
%%%
%%%
\noindent
As shown in  Fig.~\ref{fig:noise-neg},  there is a range of $\Omega$ in which  entanglement cannot be detected using our setup. For $\Omega\sigma=0.5$ (small values of $\Omega$), entanglement is not harvestable in the whole parameter region of $(t_0, x_A)$ except the very small region near the mirror. For $\Omega\sigma=1$, entanglement can be harvestable in the whole region of $(t_0,x_A)$ but the amount of entanglement is reduced in the region with Hawking radiation. For $\Omega\sigma=3$, Hawking radiation reduces the amount of entanglement and results in entanglement death.

For a larger $\kappa$ (higher temperature), the harvestable range of $\Omega$ decreases, and above a critical value of $\kappa$, entanglement cannot be detected. This is similar to the result of \cite{Simidzija-Martin-Martinez-2018}, who considered  entanglement-harvesting in a thermal bath and investigated  $\Omega$-dependence of negativity. 

%%%%%%%%%%%%%%%%%%%%%%%%%%%%%%%%%%%%%%%%%%%%%%%%%%%%%%
\newpage
\subsection{Second-order coherence and entanglement}
\begin{figure}[H]
  \centering
  \includegraphics[width=1\linewidth]{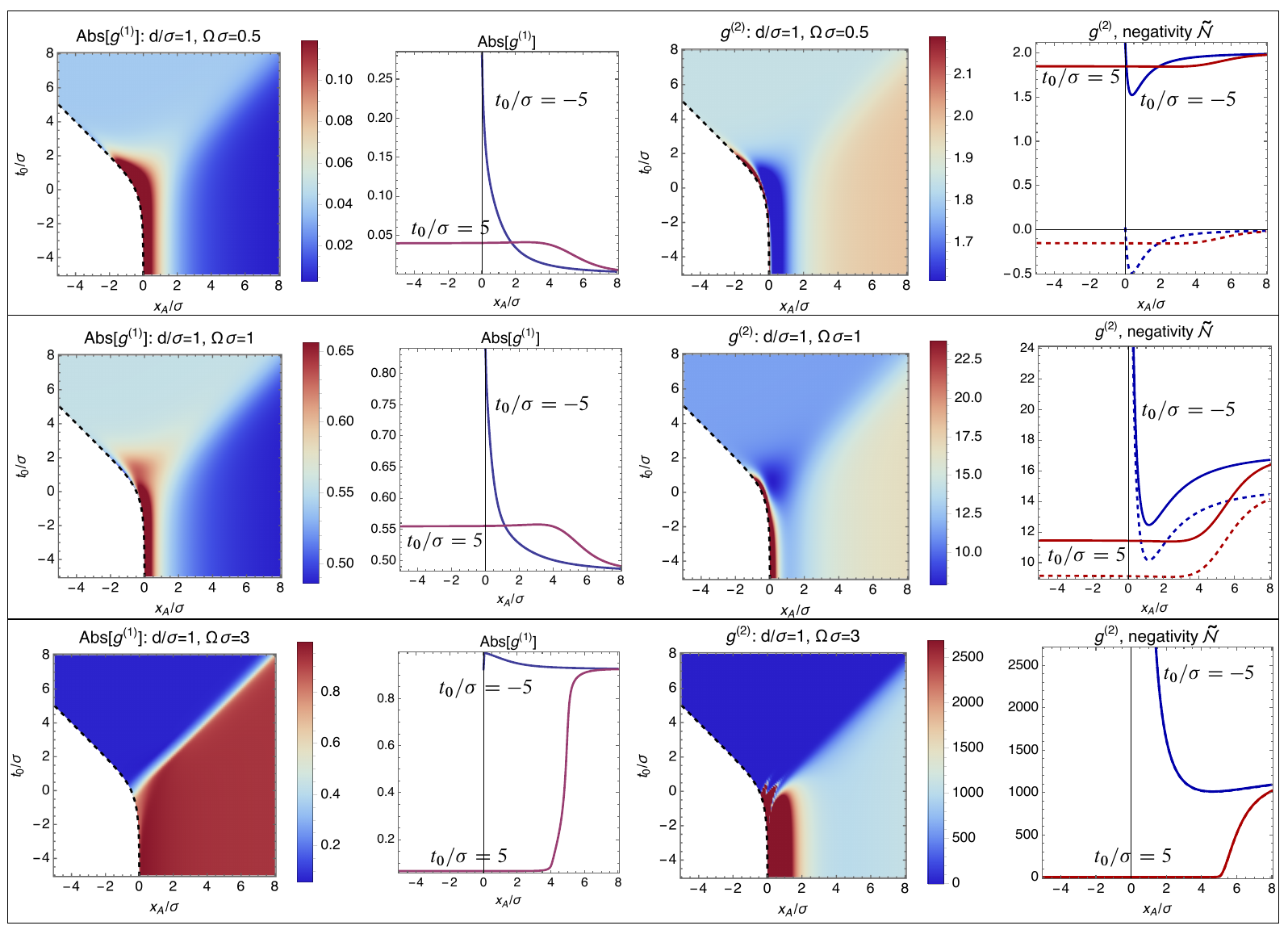}
  \caption{$g^{(1)}$ and $g^{(2)}$  with $\Omega\sigma = 0.5, 1, 3$ (from the top panels to the bottom panels). Other parameters are $\kappa/\sigma=1, d/\sigma=1$. In line plots of $g^{(2)}$ and negativity, solid lines represent $g^{(2)}$ and dashed lines represent $\widetilde{\mathcal{N}}$.}
  \label{fig:g11g22}
\end{figure}

Using the saddle-point approximation, we calculate the second-order coherence in the moving mirror system. We demonstrate the behavior of $g^{(1)}$ and $g^{(2)}$ in Fig.~\ref{fig:g11g22} using the same parameters as those adopted in Fig.~\ref{fig:noise-neg}. 
%%%

In the upper panels of Fig.~\ref{fig:g11g22}, we  show $(t_0,x_A)$-dependence of  $g^{(1)}$ and $g^{(2)}$ with $\Omega\sigma=0.5$. In the region with Hawking radiation, the first-order coherence is larger than that of the vacuum region without Hawking radiation. Thus, Hawking radiation enhances the first-order coherence for $\Omega\sigma=0.5$. The second-order coherence becomes smaller by Hawking radiation (see behavior of $g^{(2)}$ at $t_0/\sigma=5$). $g^{(2)}$ is reduced by Hawking radiation, its value is smaller than two, and the two detectors are separable. 
%%%
%%%
\noindent
In the bottom panels of Fig.~\ref{fig:g11g22}, we  show $(t_0,x_A)$-dependence of  $g^{(1)}$ and $g^{(2)}$ with $\Omega\sigma=3$. In this case, Hawking radiation reduces the first-order coherence (see behavior of $g^{(1)}$ at $t_0/\sigma=5$) %anno{[why? decoherence due to Hawking radiation?]}
and its value becomes very small. Related to this behavior of $g^{(1)}$, the second-order coherence becomes small due to Hawking radiation and the two detectors become separable. Actually, at $t_0/\sigma=5$, $g^{(2)}$ becomes $\sim 1$ and based on the relation \eqref{eq:g2tN} with $|g^{(1)}|\ll 1$, we have $\widetilde{\mathcal{N}}\approx -1$ and the two detectors are separable.

\begin{figure}[H]
    \centering
    \includegraphics[width=0.75\linewidth]{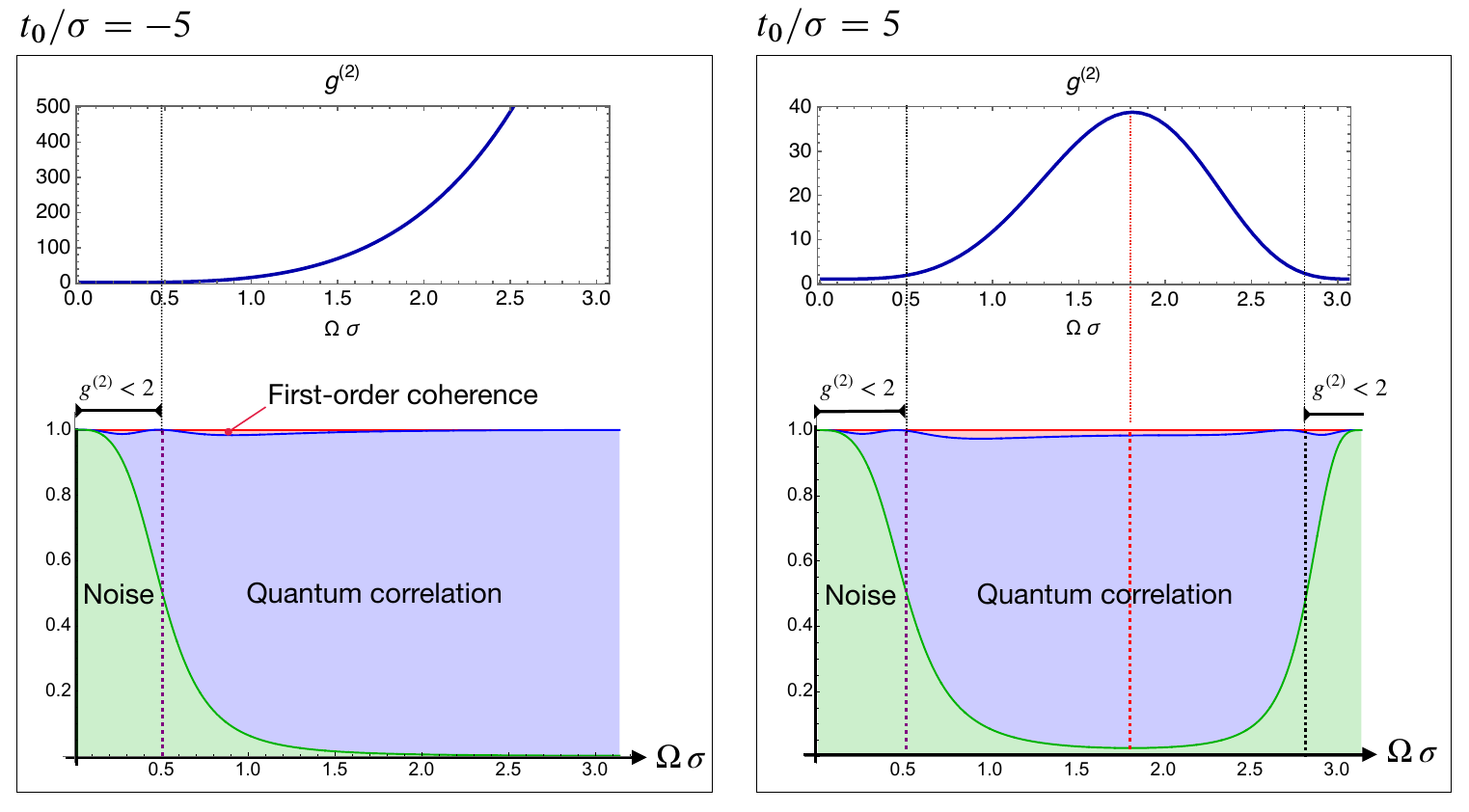}
    \caption{The breakdown of $g^{(2)}$ based on the formula \eqref{eq:g22} which shows the contribution of noise, the first-order coherence and the quantum correlation. Parameters are $\kappa/\sigma = 1.0, x_A/\sigma = 4, d/\sigma = 1$. The left panel is for $t_0/\sigma = -5$ before emission of Hawking radiation.  The right panel is for $t_0/\sigma = 5$, when  the emission of Hawking radiation from the mirror becomes stationary. In $g^{(2)}<2$ regions, two detectors are separable (Eq.~\eqref{eq:tnless0}).}
    \label{fig: ratio}
\end{figure}

Now, let us investigate the $\Omega$-dependence of the second-order coherence in more detail from the perspectives of noise and entanglement. Figure \ref{fig: ratio} shows the $\Omega$-dependence of $g^{(2)}$ and a detailed breakdown of $g^{(2)}$ according to the formula~\eqref{eq:g22} before and after the emission of Hawking radiation:
%%%
\begin{equation}
\text{Noise}:E_AE_B\qquad\text{First-order coherence}: E_{AB}\qquad\text{Quantum correlation}:|X|^2
\end{equation}
%%%
The sum of these quantities  yields $E_AE_B\times g^{(2)}$.
%%%
%%%
\noindent 
The left panels in Fig.~\ref{fig: ratio} shows the breakdown of $g^{(2)}$ at $t_0/\sigma=-5$ before emission of Hawking radiation.  The right panel shows the breakdown of $g^{(2)}$  when Hawking radiation is  emitted by the mirror.  In  both cases, the  contribution of the first-order coherence to $g^{(2)}$ is negligible compared to   other contributions.  Thus, the first-order coherence has no effect on entanglement-harvesting for this set of parameters. This was expected, as shown in Fig.~\ref{fig:noise-neg}.  At the early time $t_0/\sigma=-5$, the amount of noise $E_AE_B$ decreases  as $\Omega\sigma$ increases.  The noise is dominant  in the small $\Omega\sigma$ region.  This noise originates from vacuum fluctuations.  For $t_0/\sigma=5$,  the  noise contribution reaches a  minimum at approximately  $\Omega\sigma\sim 1.8$.  This breakdown behavior  differs  from that for $t_0/\sigma = -5$ case. The difference between these two  behaviors is due to the existence of thermal Hawking radiation. The increase of noise at $\Omega\sigma > 1.8$ is  due to Hawking radiation. Similar behavior can also be observed  for negativity in a  thermal bath \cite{Simidzija-Martin-Martinez-2018} (see also Fig.~\ref{fig:EAneg} which took into account of the thermal contribution of the Wightman function $D_1$).

%%%%%%%%%%%%%%%%%%%%%%%%%%%%%%%%%%%%%%%%%%%%%%%%%%%%%%%%%%%%%%%%
%%%%%%%%%%%%%%%%%%%%%%%%%%%%%%%%%%%%%%%%%%%%%%%%%%%%%%%%%%%%%%%%
%%%%%%%%%%%%%%%%%%%%%%%%%%%%%%%%%%%%%%%%%%%%%%%%%%%%%%%%%%%%%%%%

\section{Conclusion}\label{sec. conclusion}

 In this study, we investigate the properties of the second-order coherence of Hawking radiation in a moving-mirror model.
 We confirm that it can also serve as a measure of quantum entanglement in quantum field theory.  
 We also investigated the effects of vacuum fluctuations and Hawking radiation (thermal noise) on the entanglement-harvesting protocol based on the intensity correlation $g^{(2)}$.
 From the general property of the intensity correlation, we can conclude that for $g^{(2)}<2$, the two-qubit detector state is separable and for $3<g^{(2)}$, the state is entangled and entanglement-harvesting succeeds. In the moving-mirror system investigated in this study, the first-order coherence is small, and the second-order coherence becomes a good indicator of entanglement because the relation $g^{(2)}\approx 2+\tilde N$ holds.

The second-order coherence includes effect of purely quantum correlation that originates from the Bell non-locality. As shown in the previous study~\cite{Numbu-Ohsumi-2011}, a sufficient condition for violation of the Bell-CHSH inequality \cite{PhysRevLett.23.880} is given by 
%%%
\begin{equation}
 |X|^4>16X_4E_AE_B.
\end{equation}
%%%
Using the relation $g^{(2)}\approx\widetilde N+2$, the condition of the Bell non-locality is represented as a condition for the second-order coherence 
$ ( g^{(2)})^2-18g^{(2)}+1>0$ and this inequality yields $g^{(2)}>9+4\sqrt{5}=17.9443$. For this range of $g^{(2)}$, we know that the two-qubit state is entangled from \eqref{eq:tngtr0}. Thus, it is possible to obtain the property of Bell non-locality from the second-order coherence in the moving mirror system.

 As a further development of this study, we would like to consider the behavior of $g^{(2)}$ when the accelerated mirror subsequently decelerates to stand still. Such a mirror trajectory corresponds to the scenario of completely evaporating BHs, and we expect that the purification partner of the Hawking radiation will return to our observable region, and the unitarity of the system will be recovered~\cite{Osawa-Lin-Nambu-Hotta-2024}. 
If $g^{(2)}$ is evaluated for this type of mirror trajectory, we expect a different behavior of $g^{(2)}$ investigated in this study due to the different asymptotic null structure of spacetime that leads to the different entanglement behavior measured by an asymptotic observer. Therefore, evaluating $g^{(2)}$ for different mirror trajectories would be valuable to further understanding the structure of the entanglement between Hawking radiation and its partner from the perspective of the second-order coherence.

%%%%%%%%%%%%%%%%%%%%%%%%%%%%%%%%%%%%%%%%%%%%%%%%%%%%%%%%%%%%%%%%
%%%%%%%%%%%%%%%%%%%%%%%%%%%%%%%%%%%%%%%%%%%%%%%%%%%%%%%%%%%%%%%%
%%%%%%%%%%%%%%%%%%%%%%%%%%%%%%%%%%%%%%%%%%%%%%%%%%%%%%%%%%%%%%%%

\begin{acknowledgements}
MT was supported by the ``THERS Make New Standards Program for Next Generation Researchers'' program at Nagoya University (JST SPRING, Grant Number JPMJSP2125).
YN was supported by JSPS KAKENHI (Grant No.~JP22H05257) and MEXT KAKENHI Grant-in-Aid for Transformative Research Areas A ``Extreme Universe"(Grant No.~24H00956).
\end{acknowledgements}

\bibliography{ref} 

\end{document}